\documentclass[12pt,epsf]{article}

\usepackage{epsfig}
\usepackage{amssymb}
\usepackage{graphicx}
\usepackage{color}
\usepackage{subfigure}

\begin{document}

\baselineskip 6mm
\renewcommand{\thefootnote}{\fnsymbol{footnote}}


\newcommand{\nc}{\newcommand}
\newcommand{\rnc}{\renewcommand}

\headheight=0truein
\headsep=0truein
\topmargin=0truein
\oddsidemargin=0truein
\evensidemargin=0truein
\textheight=9.5truein
\textwidth=6.5truein

\rnc{\baselinestretch}{1.24}    
\setlength{\jot}{6pt}       
\rnc{\arraystretch}{1.24}   



\newcommand{\tcb}{\textcolor{blue}}
\newcommand{\tcr}{\textcolor{red}}
\newcommand{\tcg}{\textcolor{green}}


\def\be{\begin{equation}}
\def\ee{\end{equation}}
\def\ba{\begin{array}}
\def\ea{\end{array}}
\def\bea{\begin{eqnarray}}
\def\eea{\end{eqnarray}}
\def\nn{\nonumber\\}


\def\ct{\cite}
\def\la{\label}
\def\eq#1{Eq. (\ref{#1})}


\def\a{\alpha}
\def\b{\beta}
\def\g{\gamma}
\def\G{\Gamma}
\def\d{\delta}
\def\D{\Delta}
\def\ep{\epsilon}
\def\e{\eta}
\def\ph{\phi}
\def\Ph{\Phi}
\def\ps{\psi}
\def\Ps{\Psi}
\def\k{\kappa}
\def\l{\lambda}
\def\L{\Lambda}
\def\m{\mu}
\def\n{\nu}
\def\th{\theta}
\def\Th{\Theta}
\def\r{\rho}
\def\s{\sigma}
\def\S{\Sigma}
\def\ta{\tau}
\def\o{\omega}
\def\O{\Omega}
\def\pr{\prime}


\def\half{\frac{1}{2}}

\def\goto{\rightarrow}

\def\na{\nabla}
\def\grad{\nabla}
\def\curl{\nabla\times}
\def\div{\nabla\cdot}
\def\pa{\partial}

\def\ll{\left\langle}
\def\rr{\right\rangle}
\def\lb{\left[}
\def\lc{\left\{}
\def\ls{\left(}
\def\ln{\left.}
\def\rn{\right.}
\def\rb{\right]}
\def\rc{\right\}}
\def\rs{\right)}

\def\vac#1{\mid #1 \rangle}


\def\td#1{\tilde{#1}}
\def\check{ \maltese {\bf Check!}}


\def\Tr{{\rm Tr}\,}
\def\det{{\rm det}}


\def\bc#1{\nnindent {\bf $\bullet$ #1} \\ }
\def\ch {$<Check!>$ }
\def\ss {\vspace{1.5cm}}

\begin{titlepage}
\hfill\parbox{5cm} { }

\vspace{25mm}

\begin{center}
{\Large \bf A Dual Geometry of the Hadron in Dense Matter}
\vskip 1. cm
  {Bum-Hoon Lee$^{ab}$\footnote{e-mail : bhl@sogang.ac.kr},
  Chanyong Park$^b$\footnote{e-mail : cyong21@sogang.ac.kr} and
  Sang-Jin Sin$^c$\footnote{e-mail : sjsin@hanyang.ac.kr}}

\vskip 0.5cm

{\it
$^a\,$ Department of Physics, Sogang University, Seoul 121-742, Korea \\
$^b\,$ Center for Quantum Spacetime (CQUeST), Sogang University, Seoul 121-742, Korea \\
$^c\,$ Department of physics,  Hanyang University, \\
 Seoul 133-791, Korea}\\

\end{center}

\thispagestyle{empty}

\vskip2cm


\centerline{\bf ABSTRACT} \vskip 4mm

\vspace{1cm}
We identify the dual geometry of the hadron phase of dense nuclear matter  and investigate the confinement/deconfinement
phase transition.
We suggest that the low temperature phase
of the RN black hole with the full backreaction
of the bulk gauge field is described by the zero mass limit of the
RN black hole with hard wall.
We  calculated the density dependence of  critical temperature
and found that the phase diagram closes.
We also study   the density dependence of the $\r$ meson mass.
\vspace{2cm}


\end{titlepage}

\renewcommand{\thefootnote}{\arabic{footnote}}
\setcounter{footnote}{0}

\tableofcontents

\section{Introduction}
Recently, related to the RHIC and LHC experiments,
understanding strongly interacting QCD
 is requesting much attention.  Although a powerful method for this subject, the lattice QCD,  is being developed,  when it comes to
the dense matter problem, lattice calculation has difficulty and
not much result is produced so far.
The AdS/CFT correspondence\cite{Maldacena:1997re} in the string theory, can shed many
aspect of hadron theory\cite{Sakai:2004cn,Kim:2006gp,Horigome:2006xu,Nakamura:2006xk,
Kobayashi:2006sb,Parnachev:2006ev,Domokos:2007kt,Kim:2007em,Sin:2007ze}
as well as in strongly interacting quark gluon
plasma\cite{Shuryak:2004cy,Policastro:2001yc,Sin:2004yx,Nastase:2005rp,
Janik:2005zt,Herzog:2006gh}. The theory
can easily accommodate the  dense matter problem at least for deconfined phase
\cite{Kim:2007zm,Horigome:2006xu,Nakamura:2006xk,
Kobayashi:2006sb}. However, for the
the hadron phase, the status is not very clear since
even the phase diagram is qualitatively different from that  of the real QCD
\cite{Horigome:2006xu,Bergman:2007wp}.
This may be traced back to the probe approach of the dense matter and one expect that
if one fully account the back reaction of the gravity to the dense matter,
one might overcome the situation.

In fact, in the previous work\cite{Sin:2007ze}, it was suggested that if one considers bulk filling branes, one can easily account the full back reaction and the phase diagram actually closes.
However, in that work,  the back reacted geometry  of  the hadronic phase  was not fully identified but was approximately treated as the thermal ads with electric potential.

In this paper we consider the zero black hole mass limit of the charged
black hole with hard wall installed as the low energy pair of the charged black hole. Although it has the naked singularity at the origin
it is hidden in the wall and we do not find any physical difficulty.
With this identification,
density dependence of the physical quantities  can be easily calculated. As a first example, we calculated how meson spectrum
depends on the density of the baryonic medium.

The rest of paper follows: In the section 2, we will briefly review the dual geometry for the quark-gluon plasma   and then propose what
is the dual geometry for the hadronic  phase. In the section 3,   we will investigate the Hawking-Page transition.  We will first study the fixed chemical potential case and then consider  the
fixed density problem.  In the section 4,
we will investigate the mass of the excited vector mesons in dense medium. In the section 5, we will summarize our results and discuss some future works.

\section{Dual geometry for QCD with quark matters}

In AdS/CFT correspondence, the boundary value of the bulk gauge field
is coupled to the dual operator in the QCD side, which is the quark current. Furthermore the boundary value of the time-component gauge field is the chemical potential its dual operator is the quark number density. Our main interest here is to see how the critical temperature of the phase transition depends on the quark chemical potential. To describe the region for the high chemical potential, we need to consider the back reaction
of the bulk gauge field in the dual geometry. Here, we will investigate
the asymptotically AdS geometry dual to QCD with hadronic matters.

We first review the gravity theory  in the
Mikowskian signature with introducing our conventions.
The gravity action describing the five-dimensional asymptotic AdS space
with the gauge field is given by
\be	\la{Maction}
S_M = \int d^5 x \sqrt{-G} \lb \frac{1}{2 \k^2} \ls   {\cal R} -  2 \L \rs
- \frac{1}{4g^2} F_{MN} F^{MN} \rb,
\ee
where $2 \k^2$ is proportional to the five-dimensional Newton constant and $g^2$ is a
five-dimensional gauge coupling constant. In the five dimensional AdS space, the cosmological
constant is given by $\L = - 6/R^2$, where $R$ is the  radius of the AdS space.
The equations of motion of this system becomes
\bea    \la{Eeqmo}
&& {\cal R}_{MN} - \half G_{MN} {\cal R} + G_{MN} \L =  \frac{\k^2}{g^2}
\ls F_{MP} F_{N}^{P} - \frac{1}{4} G_{M N} F_{PQ} F^{P Q} \rs , \nn
&& 0 = \pa_{M} \sqrt{-G} G^{M P} G^{NQ} F_{PQ} ,
\eea
where $M,N = 0,1,\cdots,4$, $x^0=t$ and $x^4=z$.
Under the following ansatz
\bea	\la{Man}
A_0 &=& A_0 (z) , \nn
A_i &=& A_4 = 0  \ \ \ (i=1,\cdots,3) ,\nn
ds^2 &=& \frac{R^2}{z^2} \ls  - f(z) dt^2 + d x_i^2 + \frac{1}{f(z)} dz^2 \rs ,
\eea
the most general solution of \eq{Eeqmo} known as the Reissner-Nordstrom AdS
black hole (RNAdS BH) is
\bea
f(z) &=&   1 - m z^4 + q^2 z^6 , \nn
A_0 &=&   \m -Q z^2 ,
\eea
where the boundary space, denoted by $x^{\m} = \lc x^0,x^i \rc$, is located at $z=0$.
So, $\m$ is a boundary value of $A_0$ and $Q$ is related to the black
hole charge $q$
through
\be \la{denrel}
q^2 = \frac{2 \k^2}{3 g^2 R^2} Q^2.
\ee
In the AdS/QCD context \ct{Sin:2007ze}, the gravitation constant $2 \k^2$
and the five-dimensional
coupling constant $g^2$ are related to the rank of the gauge group $N_c$  and
the number of the flavor $N_f$ in QCD
\be
\frac{1}{2 \k^2} = \frac{N_c^2}{8 \pi^2 R^3} \ \ {\rm and} \ \
\frac{1}{g^2} = \frac{N_c N_f}{4 \pi^2 R} ,
\ee
so that \eq{denrel} can be rewritten as
\be
Q = \sqrt{\frac{3}{2} \frac{N_c}{N_f}} \ \ q .
\ee

\subsection{Dual geometry of the quark-gluon plasma}

For investigating the Hawking-Page transition dual to the C/D phase transition in QCD,
it is more convenient to consider the Euclidean version.
Here, we summarize the Euclidean RNAdS BH shortly, which corresponds to the deconfinement
phase described by the quark-gluon plasma.

By the Wick rotation $t \to - i \ta$, the Euclidean version of the previous action
in \eq{Maction} reads
\be \la{Eact}
S = \int d^5 x \sqrt{G} \lb \frac{1}{2 \k^2} \ls  - {\cal R} + 2 \L \rs
+ \frac{1}{4g^2} F_{MN} F^{MN} \rb ,
\ee
where $G_{MN}$ is the Euclidean metric ansatz
\bea
ds^2 &=& \frac{R^2}{z^2} \ls  f(z) d \ta^2 + d \vec{x}^2 + \frac{1}{f(z)} dz^2 \rs .
\eea
The equations of motion for this system become
\bea    \la{eqmo}
&& {\cal R}_{MN} - \half G_{MN} {\cal R} + G_{MN} \L =  \frac{\k^2}{g^2}
\ls F_{MP} F_{N}^{P} - \frac{1}{4} G_{MN} F_{PQ} F^{PQ} \rs , \nn
&& 0 = \pa_{M} \sqrt{G} G^{MP} G^{NQ} F_{PQ} .
\eea
Under the following ansatz
\bea
A_{\ta} &=& A(z) , \nn
A_i &=& A_ 4 = 0   ,
\eea
the most general solution is nothing but the Euclidean RNAdS BH,
so that the metric factor $f(z)$ is the same as one in the Minkowski version.
Finally, the metric solution is
\be \la{rnbh}
ds^2 = \frac{R^2}{z^2} \ls  ( 1 - m z^4 + q^2 z^6 ) d \ta^2 + d \vec{x}^{ 2}
+ \frac{1}{1 - m z^4 + q^2 z^6 } dz^2 \rs .
\ee
From now on, we consider the Euclidean version only.
Because $F_{zt}=-F_{tz}= \pa_z A_t$, the Maxwell equation can be reduced to the
simple form
\be
0 = \pa_z \ls \frac{R}{z} \ \pa_z A(z) \rs ,
\ee
and the solution is given by
\be \la{solF}
A(z) = i \ls \m - Q z^2 \rs .
\ee
Note that the imaginary number, $i$, in the above
is very important
to satisfy the Einstein equation in \eq{eqmo}, which
naturally appears  due to the Wick rotation.

From the metric in \eq{rnbh}, the outer horizon denoted by $r_+$ should satisfy
\be
0 = f(z_+) = 1 - m z_+^4 + q^2 z_+^6 .
\ee
Using the above, we can replace the black hole mass $m$ with a function of the outer horizon
$z_+$ and the black hole charge $q$
\be \la{mr}
m = \frac{1}{z_+^4} + q^2 z_+^2  ,
\ee
which is useful for the later convenience.
The Hawking temperature of the RNAdS BH is given by
\be 	\la{rntemp}
T_{RN} = \frac{1}{\pi z_+} \ls 1 - \half q^2 z_+^6 \rs .
\ee
For the norm of $||A(z)|| \equiv g^{\ta \ta} A_\ta A_{\ta}$ at the black hole horizon
to be regular, we should impose
the Dirichlet boundary condition $A(z_+) = 0$ \ct{Sin:2007ze,Horigome:2006xu,Nakamura:2006xk,
Hawking:1995ap,Chamblin:1999tk}, which gives a relation
between $Q$ and $\m$
\be	\la{relmq}
Q^2 = \frac{\m^2}{z_+^4} .
\ee
Inserting this relation and \eq{denrel} into \eq{rntemp},
we can find
$z_+$ as a function of $\m$ and $T_{RN}$
\be \la{zpft1}
z_+ = \frac{3 g^2 R^2}{2 \k^2 \m^2}
\ls \sqrt{\pi^2 T_{RN}^2 + \frac{4 \k^2 \m^2}{3 g^2 R^2}} - \pi T_{RN} \rs .
\ee

To describe a system having the fixed chemical potential, we should impose
the Dirichlet boundary condition, $A(0) = i \mu$, at the UV cut-off $z=\ep$
where $\ep$ is very small. Then, the on-shell action becomes
\be
S^D_{RN} = \frac{V_3 R^3}{\k^2} \frac{1}{T_{RN}}
\ls \frac{1}{\ep^4} - \frac{1}{z_+^4} - \frac{2 \k^2}{3 g^2 R^2} \frac{\m^2}{z_+^2} \rs ,
\ee
where $V_3$ corresponds to the spatial volume of the boundary space. In
the above action, the superscript $D$ and the subscript $RN$
imply the Dirichlet boundary condition at the UV cut-off and the RNAdS BH, respectively.
Since this action has a divergent term when $\ep \to 0$, we should renormalize it.
For the renomalization, we use the background subtraction method, in which the on-shell
action for the AdS space is subtracted from $S^D_{RN}$. The on-shell action for
the AdS space is
\bea    \la{sac}
S_{AdS} &=& \frac{V_3}{2 \k^2} \int_0^{\b} d \ta \int_{\ep}^{\infty} dz \
\sqrt{G}  \ls - {\cal R} + 2 \L \rs \nn
&=& \frac{V_3 R^3}{\k^2}  \frac{\b}{\ep^4} .
\eea
After identifying the circumference of the RNAdS BH and the AdS space at the boundary, we
can rewrite $\b$ as
\be
\b = \frac{1}{T_{RN}} \ls 1 - \half m \ep^4 + {\cal O} (\ep^6) \rs .
\ee
Therefore, the on-shell action of the AdS space becomes
\be
S_{AdS} = \frac{V_3 R^3}{\k^2} \frac{1}{T_{RN}} \ls \frac{1}{\ep^4}
- \frac{1}{2 z_+^4} - \frac{\k^2}{3 g^2 R^2} \frac{\m^2}{z_+^2}  \rs ,
\ee
where \eq{denrel}, \eq{mr} and \eq{relmq} are used. As a result, the renormalized action
of the RNAdS BH is given by
\bea
\bar{S}^D_{RN} &=& S^D_{RN} - S_{AdS} \nn
&=& - \frac{V_3 R^3}{\k^2} \frac{1}{T_{RN}}
\ls  \frac{1}{2 z_+^4} + \frac{\k^2}{3 g^2 R^2} \frac{\m^2}{z_+^2} \rs ,
\eea
and the grand potential becomes
\bea
\O_{RN} &=& \bar{S}^D_{RN} T_{RN}  \nn
&=& - \frac{V_3 R^3}{\k^2}
\ls  \frac{1}{2 z_+^4} + \frac{\k^2}{3 g^2 R^2} \frac{\m^2}{z_+^2} \rs .
\eea

For describing the dependence of the particle number in this system, we should consider the
free energy $F$ obtained by the Legendre transformation of the grand potential.
In the thermodynamics, it is given by
\be
F = \O + \m N ,
\ee
where $N = - \frac{\pa \O}{\pa \m}$. Using the fact that $z_+$ can be represented
as a function of $\m$ and $T$ like \eq{zpft1}, $N$ is given by
\be
N =  \frac{2   R}{g^2}  Q V_3 ,
\ee
where
the particle number $N$ is proportional to the quark number density $Q$.
As a result, the free energy becomes
\be \la{freeen}
F = \O + \frac{2 R}{g^2} \m Q V_3.
\ee

Interestingly, this result can be reobtained from the action \eq{Eact} with the
different boundary condition at the UV cut-off.
For changing the Dirichlet boundary condition into the Neumann boundary condition,
we should add a boundary term to fix $Q$. Then, the
renormalized action is given by
\be
\bar{S}_{RN}^N = \bar{S}_{RN}^D + S_{b} ,
\ee
where the superscript $N$ implies the Neumann boundary condition at the UV cut-off
and the boundary action $S_b$ is given by
\be
S_b =  \frac{1}{g^2} \int_{\pa {\cal M}} d^4 x \sqrt{G^{(4)}} \  n^{M} A^{N} F_{MN} .
\ee
In the above, the unit normal vector is given by
$n^{M} = \lc 0,0,0,0, \frac{z}{R} \sqrt{f(z)} \rc$
and $G^{(4)} = \frac{R^8}{z^8} f(z)$ is a determinant of the four-dimensional boundary
metric. Using the solution for the bulk gauge field in \eq{solF}, the boundary term
becomes
\be \la{bndact}
S_b = \frac{V_3 }{T_{RN} } \frac{2 R}{g^2} \m Q .
\ee
So, the free energy reads
\be
F = \bar{S}_{RN}^N T_{RN} = \O + \frac{2R}{g^2} \m Q V_3 ,
\ee
which is the same as one obtained from the thermodynamics in \eq{freeen}.
From these results,
we may conclude that the five-dimensional bulk action with the Dirichlet
or Neumann boundary condition at the UV-cut off
corresponds to the grand potential
or free energy of the dual QCD, respectively.

\subsection{Dual geometry of the hadronic phase}

In QCD, it is well known that there exist the hadronic or confinement phase
at the low temperature. In the absence of quark matters, the Schwarzschild AdS black hole
(SAdS BH)
corresponds to the deconfinement phase. At the low temperature,
the dual geometry for the confinement
phase is given by the thermal AdS with the IR cut-off, which is needed
to explain the confining behavior \ct{Herzog:2006ra}. What is the dual geometry
describing the hadronic phase, in other words the confinement phase with quark matters?
As shown in the previous section, we should include the bulk gauge field to explain
the quark matters. Therefore, the geometry corresponding to the hadronic phase
has to be a deformed AdS including the backreaction of the bulk gauge field,
which is not a black hole. The metric we find out to answer the above question, is
\be \la{ltmet}
ds^2 = \frac{R^2}{z^2} \ls ( 1 + q^2 z^6 ) d \ta^2 + d \vec{x}^2
+ \frac{1}{1 + q^2 z^6 } dz^2 \rs ,
\ee
which together with \eq{solF} satisfies the Einstein and Maxwell equation and becomes
a AdS space asymptotically. This solution can be also easily obtained from the RNAdS BH
by taking $m = 0$. For the convenience, we call this solution as a
thermal charged AdS (tcAdS) solution.
Especially, in the case of $q=0$, the tcAdS and
RNAdS BH are reduced to the thermal AdS (tAdS) and SAdS BH, respectively.
Here, our proposition is that
the tcAdS having the IR cut-off $z_{IR}$
is the dual geometry corresponding
to the hadronic phase.

Now, we study the thermodynamics of the tcAdS with the fixed chemical potential, for
which the Dirichlet boundary
condition is needed at the UV cut-off. At the IR cut-off
we impose another Dirichlet boundary condition
\be
A(z_{IR}) = i \a \m ,
\ee
where $\a$ is a arbitrary constant and will be determined later.
This IR boundary condition together with \eq{solF} gives a relation between $\m$ and $Q$
\be
Q =  \frac{(1-\a) \m}{z_{IR}^2 } .
\ee
Using this, the on-shell action of the tcAdS is given by
\be
S^D_{tc} = \frac{V_3 R^3}{\k^2} \frac{1}{T_{tc}}
\ls \frac{1}{\ep^4} - \frac{1}{z_{IR}^4} - \frac{2 \k^2}{3 g^2 R^2}
\frac{(1-\a)^2 \m^2}{z_{IR}^2} \rs ,
\ee
where the subscript $tc$ means the tcAdS. To renormalize this action, we subtract
the on-shell action of the AdS space in \eq{sac}, with the identification between
the circumferences of two backgrounds at the UV cut-off
\be
\b = \frac{1}{T_{tc}} \ls 1 + {\cal O} (\ep^6) \rs .
\ee
Then, in the limit of $\ep \to 0$ the renormalized action
of the tcAdS becomes
\bea
\bar{S}^D_{tc} &=& - \frac{V_3 R^3}{\k^2} \frac{1}{T_{tc}}
\ls \frac{1}{z_{IR}^4} + \frac{2 \k^2}{3 g^2 R^2}
\frac{(1-\a)^2 \m^2}{z_{IR}^2} \rs .
\eea
From this, the particle number $N$ is given by
\be
N = \frac{2}{3} (1-\a) \frac{2R}{g^2} Q V_3 .
\ee
Since the action with the Neumann condition should be a free energy
as previously mentioned, $\m N = S_b T_{tc}$ should be satisfied.
Using the fact that the boundary action $S_b$ of the tcAdS is given by
\be
S_b = \frac{\m }{T_{tc}} \frac{2 R}{g^2} Q V_3 ,
\ee
$\a$ should becomes $-1/2$ for the consistency. As a result,
the renormalized action becomes
\be
\bar{S}^D_{tc} = - \frac{V_3 R^3}{\k^2} \frac{1}{T_{tc}}
\ls \frac{1}{z_{IR}^4} + \frac{3 \k^2}{2 g^2 R^2}
\frac{ \m^2}{z_{IR}^2} \rs ,
\ee
with the following relation
\be
\m = \frac{2}{3} Q z_{IR}^2 .
\ee

\section{Confinement/deconfinement phase transition}

In QCD, there exists the C/D phase transition, which
is dual to the Hawking-Page transition
in the gravity theory side. So it is an interesting question to ask how the
C/D phase transition depends on the quark matters.
As mentioned previously, we should introduce the IR cut-off to describe the
confining behavior of the hadronic phase, which is called the hard wall
model. There were many interesting works to explain QCD depending on quark matters.
In our previous work \ct{Kim:2007em}, we studied the dependence of the quark
(number) density in
the C/D phase transition by considering
the gauge field fluctuation on the tAdS and SAdS BH backgroud. Unfortunately,
its result is valid only in the regime of the low quark number density so that
it can not explain the dependence
of the high chemical potential and quark density. Anyway, the crucial point of
the tcAdS and RNAdS BH backgrounds is that
since they include the full backreaction of the gauge field, we can investigate
the dependence of quark matters even in the high chemical potential or quark
density regime using this model.

\subsection{Fixed quark chemical potential}

For the fixed chemical potential, to describe the Hawking-Page transition we calculate
the difference between the on-shell actions of two backgrounds, with the Dirichlet
boundary condition at the UV cut-off,
\be
\D S = S^D_{RN} - S^D_{tc} ,
\ee
with
\bea
S^D_{RN} &=& \frac{V_3 R^3}{\k^2} \frac{1}{T_{RN}}
\ls \frac{1}{\ep^4} - \frac{1}{z_+^4} - \frac{2 \k^2}{3 g^2 R^2} \frac{\m^2}{z_+^2} \rs ,\nn
S^D_{tc} &=& \frac{V_3 R^3}{\k^2} \frac{1}{T_{tc}}
\ls \frac{1}{\ep^4} - \frac{1}{z_{IR}^4} - \frac{3 \k^2}{2 g^2 R^2}
\frac{\m^2}{z_{IR}^2} \rs .
\eea
After requiring the same circumference of $\ta$ at the UV cut-off, the difference
becomes
\be
\D S = \frac{V_3 R^3}{\k^2} \frac{1}{T_{RN}} \ls \frac{1}{z_{IR}^4} - \frac{1}{2 z_+^4}
+ \frac{3 \k^2}{2 g^2 R^2} \frac{\m^2}{z_{IR}^2}
- \frac{\k^2}{3 g^2 R^2} \frac{\m^2}{z_+^2} \rs ,
\ee
which is the same as one obtained from the renormalized actions in the limit of $\ep \to 0$.
The Hawking-Page transition corresponding the C/D phase transition occurs at
$\D S = 0$. Note that the C/D phase transition
occurs only in the range of $z_+ \le z_{IR}$. So we will consider the case of
$z_+ > z_{IR}$ from now on.
Suppose that $\D S$ is zero at a critical point $z_+= z_c$.
In the case of $z_+ < z_c$, $\D S$ becomes negative. So the RNAdS BH is stable, which
implies that the dual boundary theory is
described by quark-gluon plasma or the deconfinement phase.
In the Schwarzschild black hole case,
since there is no
black hole charge, the dual theory is described by the gluon only,
without including quark matters.
Anyway, for $z_c < z_+ \le z_{IR}$ the stable space is the tcAdS. Since the tcAdS
corresponds to the confinement
phase, the dual QCD describes the hadronic matters.

\begin{figure}
\vspace{2cm}
\centerline{\epsfig{file=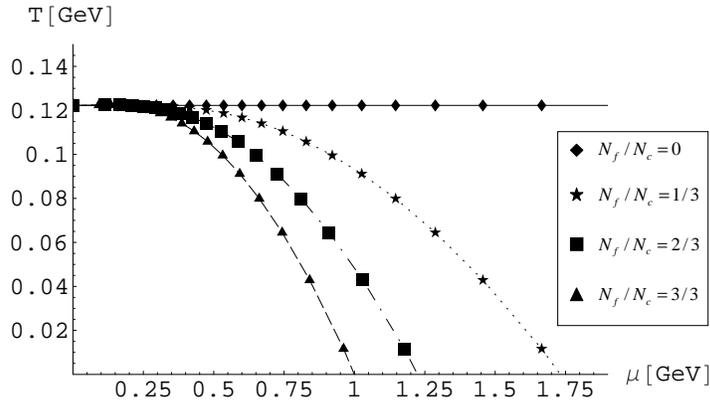,width=12cm}}
\vspace{-4.5cm}
\caption{\small The deconfinement temperature depending on the chemical
potential, which does not contain the quark mass.  }
\label{chemical}
\end{figure}


For the later convenience, we introduce dimensionless variables as the following
\bea
\td{z}_c &\equiv& \frac{z_c}{z_{IR}} , \nn
\td{\m}_c &\equiv& \m_c z_{IR} , \nn
\td{T}_c &\equiv& T_c z_{IR} ,
\eea
where the subscript $c$ means the critical values representing the C/D phase
transition point. Then, the critical
chemical potential $\td{\m}_c$
and temperature $\td{T}_c$ can be represented as functions of $\td{z}_c$
\bea
\td{\m}_c &=&
\sqrt{\frac{3 N_c}{N_f} \ \frac{(1- 2 \td{z}_c^4)}{\td{z}_c^2 (9 \td{z}_c^2 -2)}}
\ , \nn
\td{T}_c &=& \frac{1}{\pi \td{z}_c} \ls 1
- \frac{1- 2 \td{z}_c^4}{9 \td{z}_c^2 -2} \rs .
\eea
To obtain a well-defined chemical potential, the inside of the square root in
the above should become
positive. So the allowed range of $\td{z}_c$ is give by $ 0.4714 \lesssim  \td{z}_c
\lesssim 0.8409$. Though at $\td{\m}_c = 0$ the deconfinement temperature
$T_c = 122 \ {\rm MeV}$
is lower than the lattice estimation $T_c = 175 \sim 190 \ {\rm MeV}$
\ct{Karsch:2006xs,Aoki:2006br}
as noted by Herzog \ct{Herzog:2006ra},
we use $T_c = 122 \ {\rm MeV}$ and look at the qualitative behavior of
the deconfinement temperature depending on
the chemical potential.
We obtain the expected phase diagram, which is given in  the Figure 1.
Notice that the deconfinement temperature decreases more quickly
as the ratio $N_f/N_c$ goes up.
In the holographic models with probe brane approach, the phase diagram does not
close. However, in our case the phase diagram closes due to the backreaction
of the gauge field as emphasized in Ref. \ct{Sin:2007ze}. In the zero temperature
the critical value of the phase transition, $\m_c (T_c=0)$ is given by $1002 \ {\rm MeV}$
for $N_f/N_c=1$.
Note that our definition of the chemical potential does not include quark mass.
From this, we can evaluate
the the critical baryon chemical potential with mass,
\be
\bar{\m}_B = 3 {\m} + m_B \approx 4 {\rm GeV} ,
\ee
where $m_B$ is the baryon mass. For more realistic value, we should use different
normalization or other
models like the one suggested in Ref. \ct{Karch:2006pv}.

\subsection{Fixed quark number density}

For the fixed quark number density, we should add a boundary term for fixing $Q$,
which corresponds to the Neumann
boundary condition at the UV cut-off.
The on-shell actions of the RNAdS BH and the tcAdS are given by
\bea
S^N_{RN} &=&\frac{V_3 R^3}{\k^2}  \frac{1}{T_{RN}} \ls \frac{1}{\ep^4}
- \frac{1}{z_+^4}  +  \frac{4 \k^2 Q^2}{3 g^2 R^2 }  z_+^2 \rs ,\nn
S^N_{tc} &=&\frac{V_3 R^3}{\k^2}  \frac{1}{T_{tc}} \ls \frac{1}{\ep^4} - \frac{1}{z_{IR}^4}
+  \frac{2 \k^2 Q^2}{3 g^2 R^2}  z_{IR}^2 \rs .
\eea
Then, the difference between two on-shell actions becomes
\bea
\D S &\equiv& S^N_{RN} - S^N_{tc} \nn
&=& \frac{V_3 R^3}{\k^2} \frac{1}{T_{RN}} \lb \frac{1}{z_{IR}^4}
- \frac{1}{2 z_+^4} + \frac{\k^2 Q^2}{3 g^2 R^2}  \ls 5 z_+^2 - 2 z_{IR}^2 \rs  \rb .
\eea
Like the previous section, we assume that $z_c$ is a critical value of $z_+$
where the C/D phase transition
occurs. For $z_+  < z_c$, the RNAdS BH is more stable than the tcAdS. On the contrary,
the tcAdS is stable for
$z_+ > z_c$.

\begin{figure}
\vspace{2cm}
\centerline{\epsfig{file=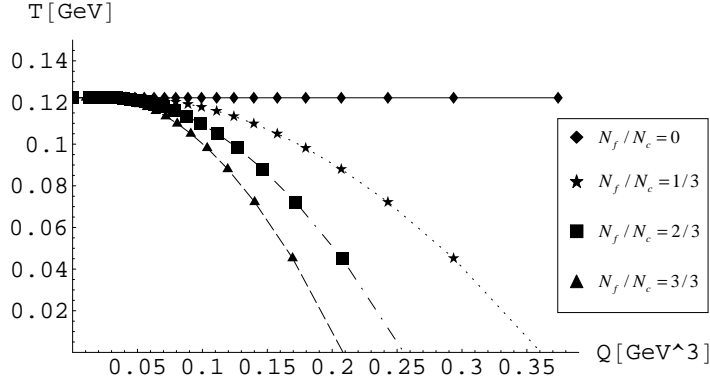,width=12cm}}
\vspace{-4.5cm}
\caption{\small The Hawking-Page transition temperature depending on the quark number
density.  }
\label{number}
\end{figure}

After introducing a new dimensionless variable, $\td{Q}_c = Q_c z_{IR}^3 $,
the dimensionless quark number density and critical temperature are given as
functions of $\td{z}_c$
\bea    \la{qt}
\td{Q}_c &=& \sqrt{\frac{3 N_c}{2 N_f}
\frac{(1- 2 \td{z}_c^4)}{\td{z}_c^4 (5 \td{z}_c^2-2)}} ,\nn
\td{T}_c &=& \frac{1}{\pi \td{z}_c} \lb  1 - \frac{\td{z}_c^2}{2} \frac{(1-2 \td{z}_c^4 )
}{ (5 \td{z}_c^2-2)} \rb.
\eea
In the first relation, $\td{Q}_c$ is well defined only in the range,
$\sqrt{\frac{2}{5}} \le \td{z}_c \le \frac{1}{2^{1/4}}$, which
is $0.6325 \lesssim \ \td{z}_c \ \lesssim 0.8409$.
Using \eq{qt}, we numerically draw the deconfinement temperature
depending on the quark density in the Figure 2. As shown in the Figure 2, the
deconfinement temperature decreases as the quark density increases. Furthermore,
when the number of flavor becomes large, the deconfinement temperature decreases
more quickly than one having the smaller number of flavor.

\section{The mass of the excited vector mesons}

\begin{figure}
\vspace{2cm}
\centerline{\epsfig{file=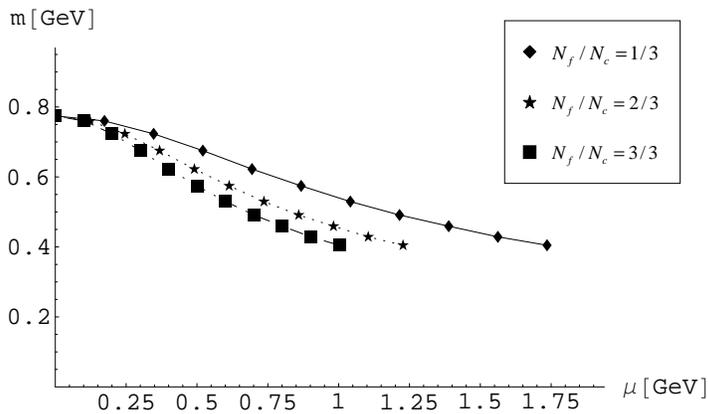,width=12cm}}
\vspace{-4.5cm}
\caption{\small The $\rho$ meson mass depending on the chemical potential.  }
\label{mesonmass}
\end{figure}

At first, we consider the $\r$ meson mass at zero temperature and finite
baryon chemical potential $\m_B$, which is given by three times as much as the quark
chemical potential, $\m_B = 3 \m_q$.
To describe this, we should consider the fluctuation of the vector field
in the tcAdS. The ansatz for the vector field fluctuation
is given by
\be
\d A_{\m} = V_{\m} (z,p) e^{i p \cdot x} .
\ee
Since the rotation symmetry in the $\ta$-$x_i$ plane is broken, we should distinguish the
$\d A_0$ with $\d A_i$. The equation of motion for $\d A_0$ is the exactly same as one
obtained in tAdS. So, from now on we will concentrate on the $\d A_i$ fluctuation
having the following equation of motion
\be \la{eqvmass}
0 = \pa_z^2 V_i - \frac{1}{z} \frac{(1-5 q^2 z^6)}{(1+q^2 z^6)} \ \pa_z V_i + m_m^2 V_i ,
\ee
where the meson mass is given by $m_m^2 = - p^2$.
After introducing the new dimensionless variables
\be
\td{V}_i = V_i z_{IR} \ \ , \ \ \ \td{z} = \frac{z}{z_{IR}} \ \ {\rm and} \ \ \
\td{m}_m = m_m z_{IR} ,
\ee
\eq{eqvmass} can be rewritten as
\be
0 = \pa_{\td{z}}^2 \td{V}_i - \frac{1}{\td{z}} \frac{(1-5 \td{q}^2 \td{z}^6)}{(1+\td{q}^2 \td{z}^6)} \
\pa_{\td{z}} \td{V}_i + \td{m}_m^2 \td{V}_i .
\ee
For $\td{q}=0$, the above equation
reproduces the result in tAdS \ct{Herzog:2006ra}.
Due to the C/D phase transition, the range of the quark chemical potential $\td{\m}$
at the zero temperature is limited from $0$ to $5.373$
for $\frac{N_f}{N_c} = \frac{1}{3}$. Similarly, $\td{\m}$ runs
from $0$ to $3.799$ for $\frac{N_f}{N_c} = \frac{2}{3}$
and from $0$ to $3.10217$ for $N_f/N_c = 1$.
Through the numerical evaluation, we find the relation between the chemical potential
and the $\rho$ meson mass in the Figure 3. As shown in the figure,
the $\r$ meson mass decreases
as the quark chemical potential increases.
Note that in the Figure 3, the right ends of the curves
implies the C/D phase transition point.
In addition, the larger the number of flavor
becomes, the more quickly the $\rho$ meson mass decreases.

\begin{center}
\begin{tabular}{|c||c|c|c|c|c|c|}
\hline
 & $ \quad \m = 0 \quad $ & $\m = 0.245$ & $\m = 0.491$ & $\m = 0.736$ & $\m = 0.982$
& $\m = 1.227$ \\
\hline \hline
mass of the 1st & $0.774$ & $0.724$ & $0.622$
& $0.530$ & $0.458$ & $0.404$\\
\hline
mass of the 2nd & $1.775$ & $1.737$ & $1.702$
& $1.704$ & $1.724$ & $1.750$\\
\hline
mass of the 3rd & $2.782$ & $2.758$ & $2.743$
& $2.747$ & $2.755$ & $2.762$\\
\hline
\end{tabular}
\end{center}
\noindent Table 1: The meson masses of the excited states depending on the chemical
potential in the GeV unit. \\

\begin{figure}
\vspace{2cm}
\centerline{\epsfig{file=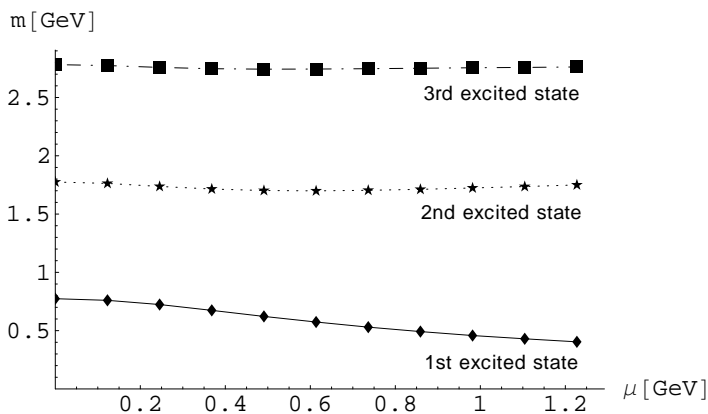,width=12cm}}
\vspace{-4.5cm}
\caption{\small For $N_f/N_c=2/3$, the chemical potential dependence of
the excited meson masses.  }
\label{mesonmass}
\end{figure}

Finally, we further calculate the masses of the higher excited meson states.
After the numerical evaluation, we draw several excited meson masses
for $\frac{N_f}{N_c} = \frac{2}{3}$ in the Figure 4 (see also Table 1. for
the numerical values of the masses in the several points of $\m$).
For the $\r$ meson, its mass decreases as the chemical potential increases.
In the higher mode cases, their masses decrease for some period of $\m$
and then increase as $\m$ increases (see Table 1).


\section{Discussion}

In this paper, we proposed the thermal charged AdS space as the
dual geometry corresponding on the hadronic phase
of QCD, which is the zero mass limit of the Reissner-Nordstrom AdS black hole with the hard wall.
This tcAdS was installed as the low temperature pair of the RNAdS black hole.
By comparing the on-shell actions of two backgrounds, we investigated the
confinement/deconfinement
phase transition depending on the chemical potential or the quark number density.
Interestingly, we found out that above the critical chemical potential
there exists only the deconfinement phase even at zero temperature and evaluated
the critical chemical potential depending on the flavor number.
In addition, using the dual geometry of the hadronic phase,
we calculated how the mass spectrum of the vector mesons
depends on the chemical potential in the baryonic medium.

In this paper, we evaluated the baryonic chemical potential ($\approx 4$ GeV), which
is too high comparing with the known result ($\approx 1$ Gev). So it is important
to know how to cure this discrepancy.
Another interesting problem is to study the chemical potential
dependence (or the quark density dependence) of other physical quantities.
We will report those results elsewhere.

\vspace{1cm}

{\bf Acknowledgement}

C. Park would like to appreciate to Youngman Kim and Kyung-il Kim and for valuable discussion.
This work was supported by the Science Research
Center Program of the Korean Science and Engineering Foundation
through the Center for Quantum SpaceTime (CQUeST) of Sogang University with grant number
R11-2005-021. The work of SJS was also supported
by KOSEF Grant R01-2007-000-10214-0.

\vspace{1cm}



\begin{thebibliography}{99}

\nc{\np}[3]{Nucl. Phys. {\bf B#1}, #2 (#3)}
\nc{\plb}[3]{Phys. Lett. {\bf B#1}, #2 (#3)}
\nc{\prl}[3]{Phys. Rev. Lett. {\bf #1}, #2 (#3)}
\nc{\prd}[3]{Phys. Rev. {\bf D#1}, #2 (#3)}
\nc{\ap}[3]{Ann. Phys. {\bf #1}, #2 (#3)}
\nc{\prep}[3]{Phys. Rep. {\bf #1}, #2 (#3)}
\nc{\ptp}[3]{Prog. Theor. Phys. {\bf #1}, #2 (#3)}
\nc{\rmp}[3]{Rev. Mod. Phys. {\bf #1}, #2 (#3)}
\nc{\cmp}[3]{Comm. Math. Phys. {\bf #1}, #2 (#3)}
\nc{\mpl}[3]{Mod. Phys. Lett. {\bf A#1}, #2 (#3)}
\nc{\cqg}[3]{Class. Quant. Grav. {\bf #1}, #2 (#3)}
\nc{\jhep}[3]{J. High Energy Phys. {\bf #1}, #2 (#3)}
\nc{\hep}[1]{{\tt hep-th/{#1}}}
\nc{\app}[3]{Ann. Phys. {\bf #1}, #2, (#3)}
\nc{\prp}[3]{Phys. Rept. {\bf #1}, #2, (#3)}
\nc{\jmp}[3]{J. Math. Phys. {\bf #1}, #2, (#3)}
\newcommand{\J}[4]{{ #1} {\bf #2} #4 (#3)}
\newcommand{\NP}{Nucl.\ Phys.}
\newcommand{\PRL}{Phys.\ Rev.\ Lett.}
\newcommand{\PL}{Phys.\ Lett.}

\bibitem{Maldacena:1997re}
  J.~M.~Maldacena,
  Adv.\ Theor.\ Math.\ Phys.\  {\bf 2}, 231 (1998)
  [Int.\ J.\ Theor.\ Phys.\  {\bf 38}, 1113 (1999)]
  [arXiv:hep-th/9711200]; \\
J. Polchinski and M. J. Strassler,
hep-th/0003136; \\
J. M. Maldacena and C. Nunez,
Phys. Rev. Lett. {\bf 86}, 588 (2001) [arXiv:hep-th/0008001]; \\
I. R. Klebanov and M. J. Strassler,
JHEP {\bf 0008}, 052 (2000)  [arXiv:hep-th/0007191].

\bibitem{Sakai:2004cn}
  T.~Sakai and S.~Sugimoto,
  Prog.\ Theor.\ Phys.\  {\bf 113}, 843 (2005)
  [arXiv:hep-th/0412141].

\bibitem{Kim:2006gp}
  K.~Y.~Kim, S.~J.~Sin and I.~Zahed,
  arXiv:hep-th/0608046.

\bibitem{Horigome:2006xu}
  N.~Horigome and Y.~Tanii,
  JHEP {\bf 0701}, 072 (2007)
  [arXiv:hep-th/0608198].

\bibitem{Nakamura:2006xk}
  S.~Nakamura, Y.~Seo, S.~J.~Sin and K.~P.~Yogendran,
  J.\ Korean Phys.\ Soc.\  {\bf 52}, 1734 (2008)
  [arXiv:hep-th/0611021].

\bibitem{Kobayashi:2006sb}
  S.~Kobayashi, D.~Mateos, S.~Matsuura, R.~C.~Myers and R.~M.~Thomson,
  JHEP {\bf 0702}, 016 (2007)
  [arXiv:hep-th/0611099].

\bibitem{Parnachev:2006ev}
  A.~Parnachev and D.~A.~Sahakyan,
  Nucl.\ Phys.\  B {\bf 768}, 177 (2007)
  [arXiv:hep-th/0610247].

\bibitem{Domokos:2007kt}
  S.~K.~Domokos and J.~A.~Harvey,
  Phys.\ Rev.\ Lett.\  {\bf 99}, 141602 (2007)
  [arXiv:0704.1604 [hep-ph]].

\bibitem{Kim:2007em}
  Y.~Kim, B.~H.~Lee, S.~Nam, C.~Park and S.~J.~Sin,
  Phys.\ Rev.\  D {\bf 76}, 086003 (2007)
  [arXiv:0706.2525 [hep-ph]].

\bibitem{Sin:2007ze}
  S.~J.~Sin,
  JHEP {\bf 0710}, 078 (2007)
  [arXiv:0707.2719 [hep-th]].

\bibitem{Shuryak:2004cy}
  E.~V.~Shuryak,
  Nucl.\ Phys.\  A {\bf 750}, 64 (2005)
  [arXiv:hep-ph/0405066]; \\
  M.~J.~Tannenbaum,
  Rept.\ Prog.\ Phys.\  {\bf 69}, 2005 (2006)
  [arXiv:nucl-ex/0603003].

\bibitem{Policastro:2001yc}
  G.~Policastro, D.~T.~Son and A.~O.~Starinets,
  Phys.\ Rev.\ Lett.\  {\bf 87}, 081601 (2001)
  [arXiv:hep-th/0104066].

\bibitem{Sin:2004yx}
  S.~J.~Sin and I.~Zahed,
  Phys.\ Lett.\  B {\bf 608}, 265 (2005)
  [arXiv:hep-th/0407215];\\
  E.~Shuryak, S.~J.~Sin and I.~Zahed,
  J.\ Korean Phys.\ Soc.\  {\bf 50}, 384 (2007)
  [arXiv:hep-th/0511199].

\bibitem{Nastase:2005rp}
  H.~Nastase,
  arXiv:hep-th/0501068.


\bibitem{Janik:2005zt}
  R.~A.~Janik and R.~B.~Peschanski,
  Phys.\ Rev.\  D {\bf 73}, 045013 (2006)
  [arXiv:hep-th/0512162]; \\
  S.~Nakamura and S.~J.~Sin,
  JHEP {\bf 0609}, 020 (2006)
  [arXiv:hep-th/0607123];\\
  S.~J.~Sin, S.~Nakamura and S.~P.~Kim,
  JHEP {\bf 0612}, 075 (2006)
  [arXiv:hep-th/0610113].

\bibitem{Herzog:2006gh}
  C.~P.~Herzog, A.~Karch, P.~Kovtun, C.~Kozcaz and L.~G.~Yaffe,
  JHEP {\bf 0607}, 013 (2006)
  [arXiv:hep-th/0605158];\\
  S.~S.~Gubser,
  Phys.\ Rev.\  D {\bf 74}, 126005 (2006)
  [arXiv:hep-th/0605182].

\bibitem{Kim:2007zm}
  K.~Y.~Kim, S.~J.~Sin and I.~Zahed,
  JHEP {\bf 0801}, 002 (2008)
  [arXiv:0708.1469 [hep-th]].

\bibitem{Bergman:2007wp}
  O.~Bergman, G.~Lifschytz and M.~Lippert,
  JHEP {\bf 0711}, 056 (2007)
  [arXiv:0708.0326 [hep-th]].

\bibitem{Hawking:1995ap}
  S.~W.~Hawking and S.~F.~Ross,
  Phys.\ Rev.\  D {\bf 52}, 5865 (1995)
  [arXiv:hep-th/9504019];\\
  R.~Emparan, C.~V.~Johnson and R.~C.~Myers,
  Phys.\ Rev.\  D {\bf 60}, 104001 (1999)
  [arXiv:hep-th/9903238].

\bibitem{Chamblin:1999tk}
  A.~Chamblin, R.~Emparan, C.~V.~Johnson and R.~C.~Myers,
  Phys.\ Rev.\  D {\bf 60}, 064018 (1999)
  [arXiv:hep-th/9902170];\\
  A.~Chamblin, R.~Emparan, C.~V.~Johnson and R.~C.~Myers,
  Phys.\ Rev.\  D {\bf 60}, 104026 (1999)
  [arXiv:hep-th/9904197].

\bibitem{Herzog:2006ra}
  C.~P.~Herzog,
  Phys.\ Rev.\ Lett.\  {\bf 98}, 091601 (2007)
  [arXiv:hep-th/0608151].

\bibitem{Karsch:2006xs}
  F.~Karsch,
  J.\ Phys.\ Conf.\ Ser.\  {\bf 46}, 122 (2006)
  [arXiv:hep-lat/0608003].

\bibitem{Aoki:2006br}
  Y.~Aoki, Z.~Fodor, S.~D.~Katz and K.~K.~Szabo,
  Phys.\ Lett.\  B {\bf 643}, 46 (2006)
  [arXiv:hep-lat/0609068].

\bibitem{Karch:2006pv}
  A.~Karch, E.~Katz, D.~T.~Son and M.~A.~Stephanov,
  Phys.\ Rev.\  D {\bf 74}, 015005 (2006)
  [arXiv:hep-ph/0602229].


\end{thebibliography}
\end{document}